\documentclass{article}
                                                                           
\usepackage{latexsym, amssymb}

\newtheorem{Proposition}{Proposition}

\newcommand{\R}{\mathbb{R}}
\newcommand{\N}{\mathbb{N}}
\newcommand{\supp}{\mathrm{supp}\,}
\renewcommand{\div}{\mathrm{div}\,}
\def\prfe{\hspace*{\fill} $\Box$

\smallskip \noindent}

\title{Global weak solutions to the Nordstr\"om-Vlasov system}
\date{}
\author{Simone Calogero\\
        Department of Mathematics, Chalmers University of Technology,\\
        S-41296 G\"oteborg, Sweden\\
        e-mail: mg026@math.chalmers.se\\       
        and\\
        Gerhard Rein\\
        University of Bayreuth, Department of Mathematics,\\
        D-95440 Bayreuth, Germany\\
        e-mail: gerhard.rein@uni-bayreuth.de\\
        phone: 0049-921-553287, fax: 0049-921-553293} 
                                                                    
\begin{document}
\maketitle
\begin{abstract}
The Nordstr\"om-Vlasov system is a Lorentz invariant model for a
self-gravitating collisionless gas. We establish suitable a-priori-bounds on
the solutions of this system, which together with energy estimates and
the smoothing effect of ``momentum averaging'' yield the existence of global weak
solutions to the corresponding initial value problem. 
In the process we improve the continuation criterion for classical
solutions which was derived recently. The weak solutions
are shown to preserve mass. 
\end{abstract}


\section{Introduction}\label{intro}
\setcounter{equation}{0}
Consider a large ensemble of particles which interact by force fields
which they create collectively and not by collisions.
Such a collisionless gas is often used
as a matter model both in plasma physics and in astrophysics. If the particles
in the gas interact by electromagnetic fields the dynamics of the ensemble 
is described by the Vlasov-Maxwell system. If the particles interact
by gravitational forces one obtains the Vlasov-Poisson system in the
non-relativistic case or the
Einstein-Vlasov system in the relativistic case.
We refer to \cite{And,Gl,Re} for background information on these systems.
In the present investigation we consider a self-gravitating collisionless 
gas where gravity is described by Nordstr\"om's scalar theory \cite{No}. 
We write the Nordstr\"om-Vlasov system in the formulation 
of \cite{CG}:
\begin{equation}\label{wave}
\partial_t^2\phi-\bigtriangleup_x\phi=-\mu,
\end{equation}
\begin{equation} \label{mudef}
\mu(t,x) = \int f(t,x,p)\,\frac{dp}{\sqrt{1+p^2}},
\end{equation}
\begin{equation} \label{vlasov}
\partial_t f + \widehat{p} \cdot \nabla_x f 
- \left[(S\phi)\,p + (1+p^2)^{-1/2} \nabla_x\phi \right]\cdot\nabla_p f
= 4 f\, S\phi .
\end{equation}
Here $t \in \R,\ x,p \in \R^3$ stand for time, position, and momentum,
$f=f(t,x,p)$, $\phi=\phi(t,x)$, and 
\[
S =\partial_t+\widehat{p}\cdot\nabla_x,\ \widehat{p} = \frac{p}{\sqrt{1+p^2}},\ p^2 = |p|^2;
\]
$S$ is the free-transport operator, and
$\widehat{p}$ 
denotes the relativistic velocity of a particle with momentum $p$.
Units are chosen such that the mass of each particle, 
the gravitational constant, and the speed of light are equal to unity.
A solution $(f,\phi)$ of this system 
is interpreted as follows: The space-time is a four-dimensional
Lorentzian manifold with a conformally flat metric which, in the coordinates 
$(t,x)$, takes the form
\[
g_{\mu\nu}=e^{2\phi} \textrm{diag}(-1,1,1,1).
\]
The particle density on the mass shell
in this metric is $e^{-4\phi}f(t,x,e^\phi p)$,
but it is more convenient to work with $f$ and $\phi$ as the dynamic
variables as long as it is kept in mind that $f$ itself is not the particle
density. More details on the derivation of this system are
given in \cite{Cal} where its steady states are investigated.

Like Nordstr\"om's theory, the system is not a physically correct 
model.  Nevertheless, there are good reasons for studying it: 
The Nordstr\"om-Vlasov system is a Lorentz invariant
model for a self-gravitating gas which has the correct Newtonian limit,
namely the Vlasov-Poisson system, cf.\ \cite{CL}. The
system is much simpler than the physically correct but notoriously difficult
Einstein-Vlasov system, and yet it captures some typical relativistic
effects, such us the propagation of gravitational waves. 
From a more mathematical point of view,
the hope is that by investigating this system one can learn more about
the whole class of non-linear partial differential equations in kinetic
theory to which this model belongs. For instance, one of the most celebrated
results in kinetic theory is the existence of global weak  
solutions for the Vlasov-Maxwell system due to R.~DiPerna and P.-L.~Lions 
\cite{DL}. It is the purpose of the present paper to investigate
this solution concept for the Nordstr\"om-Vlasov system. 
 
We explain how the paper proceeds and how global weak solutions
to the Nordstr\"om-Vlasov system are obtained. First one needs to establish
suitable a-priori-bounds for solutions. One set of such bounds derives
from conservation of energy. In addition,
for related systems like Vlasov-Poisson or
Vlasov-Maxwell the fact that $f$ is constant along characteristics
and that the characteristic flow preserves measure yields 
bounds on the $L^q$-norms of $f(t)$, $1 \leq q \leq \infty$. 
However, these properties do not hold in the present
situation. In the next section we overcome this difficulty and derive
a-priori-bounds for the  $L^q$-norms of $f(t)$. This is a necessary 
prerequisite for global existence of weak solutions, and at the same time
improves the continuation criterion for classical solutions which was
derived in \cite{CG}. It is interesting to note that the argument would 
not work if the sign in the field equation were reversed.
In the third section we turn to the construction of weak solutions.
First a suitable regularization of the system is 
introduced such that the regularized system does have global 
(classical) solutions which satisfy the same a-priori-bounds.
In \cite{DL} the Vlasov-Maxwell system was regularized by making the 
Maxwell equations parabolic. Here we smooth the right hand side of the 
field equation (\ref{wave}); an analogous approach was followed in \cite{Gl,KR,KRST}.
The resulting system is ``closer'' to the original one than in
\cite{DL}, in particular, it remains time reversible which allows
us to avoid various technical difficulties in what follows.
Along a sequence of solutions to regularized systems the a-priori-bounds
are shown to hold uniformly. In order to pass to the limit in the
non-linear term in the Vlasov equation we use the smoothing effect due to
``velocity averaging'' \cite{GLPS}, which we now prefer to call ``momentum averaging''.
As opposed to \cite{DL} we use this tool only to pass to the limit 
in the Vlasov equation and not for the moments of $f$ such as $\mu$, 
which are dealt with directly. 
This yields another simplification of the proof. 
Notice that the Vlasov equation (\ref{vlasov}) can be rewritten
in the form
\begin{equation} \label{vlasovdiv}
\partial_t f + \widehat{p} \cdot \nabla_x f = 
\nabla_p \cdot 
\left[\left((S\phi)\,p + (1+p^2)^{-1/2} \nabla_x\phi \right) f\right]
+ f\, S\phi 
\end{equation}
which is the form needed to apply ``momentum averaging''.
Once a global weak solution is
obtained its properties are of interest.  
For the related systems it is not known whether weak solutions
are unique or preserve energy, and the same is true in the present
situation. But we can show that the weak solutions preserve mass,
which is non-trivial because as opposed to the Vlasov-Maxwell system
this quantity is not just the integral of $f(t)$, cf.\ (\ref{masscons}).
For this purpose we need to exploit the relativistic nature of the system
and to assume that the initial datum for $f$ is
bounded. This assumption is not made in \cite{DL}, but
it allows us in addition to remove further technical difficulties from the proof,
such us the use of renormalized solutions. 
 
\section{Conservation laws and $L^q$-estimates}
\setcounter{equation}{0}

Formally, solutions of the Nordstr\"om-Vlasov system conserve mass and
energy:
\begin{equation}\label{masscons}
\int e^{-\phi} \int f \,dp\,dx = \mathrm{const},
\end{equation}
\begin{equation}\label{energycons}
\int\!\!\!\int f\,\sqrt{1+p^2}\,dp\,dx +\frac{1}{2}\int[(\partial_t\phi)^2+
(\nabla_x\phi)^2]\,dx = \mathrm{const}.
\end{equation}
Again formally, these conservation laws can be obtained by 
integrating their local counterparts
\begin{eqnarray}
&&
\partial_t\rho + \nabla_x\cdot j=0, \label{localmasscons}\\
&&
\partial_te + \nabla_x\cdot\mathfrak{p}=0,\label{localenergycons}
\end{eqnarray}
where
\begin{eqnarray}
\rho(t,x)
&=&
e^{-\phi}\int f(t,x,p)\,dp, \label{rhodef}\\ 
j(t,x)
&=&
e^{-\phi}\int \widehat{p}\,f(t,x,p)\,dp,\label{jdef}\\
e(t,x)
&=&
\int \sqrt{1+p^2}\,f(t,x,p)\,dp + 
\frac{1}{2}(\partial_t\phi(t,x))^2 +
\frac{1}{2}(\nabla_x\phi(t,x))^2, \label{edef}\\
\mathfrak{p}(t,x)
&=&
\int p\,f(t,x,p)\,dp -
\partial_t\phi(t,x)\nabla_x\phi(t,x). \label{pdef}
\end{eqnarray}

Let us now denote by $(f,\phi)\in C^1([0,T]\times\R^6)
\times C^2([0,T]\times\R^3)$ 
a classical solution of the Nordstr\"om-Vlasov
system on the interval $[0,T],\, T>0$, with initial data $f(0)
=f^\mathrm{in}\in C_c^1(\R^6)$, 
$\phi(0)=\phi_0^\mathrm{in}\in C^3_b (\R^3)$, $\partial_t\phi(0)
=\phi_1^\mathrm{in}\in C^2_b (\R^3)$,
with $f^\mathrm{in}\geq 0$. 
In the notation above the subscript $c$ indicates 
that the functions are compactly supported while the subscript $b$ indicates
that they are bounded together with their derivatives up to the indicated order. 
For such data a unique, classical solution
exists at least locally in time, cf.\ \cite{CG}. Assuming in addition that 
the initial data have finite energy 
the conservation laws stated above hold.

Let $X(s)=X(s,t,x,p)$, $P(s)=P(s,t,x,p)$ denote the characteristics 
of the Vlasov equation, i.e., the solutions of the characteristic
system
\begin{eqnarray*}
\frac{dx}{ds}
&=&
\widehat{p},\\
\frac{dp}{ds}
&=&
- (S\phi)\,p - (1+p^2)^{-1/2} \nabla_x\phi,
\end{eqnarray*}
satisfying $X(t)=x$, $P(t)=p$. By the Vlasov equation (\ref{vlasov})
the function $e^{-4\phi}f$ is constant 
along these curves, and $f$
can be represented as  
\begin{equation}\label{distr}
f(t,x,p)=f^\mathrm{in}(X(0),P(0))
\exp \left[4\phi(t,x)-4\phi^\mathrm{in}_0(X(0))\right].
\end{equation}
In particular, $f$ and hence also $\mu$ are non-negative. 
We write the function $\phi$ as  
\begin{equation} \label{phisplit}
\phi =\phi_{\mathrm{hom}} +\psi
\end{equation}
where $\phi_{\mathrm{hom}}$ is the solution of the homogeneous 
wave equation with initial data 
$\phi_0^\mathrm{in},\ \phi_1^\mathrm{in}$,
and $\psi$ is the solution of (\ref{wave}) with zero initial data. 
By Duhamel's principle, $\psi\leq 0$.
Hence (\ref{distr}) and (\ref{phisplit}) imply that
\begin{equation}\label{distr2}
f(t,x,p) \leq f^\mathrm{in}(X(0),P(0))
\exp \left[4 \phi_{\mathrm{hom}}(t,x) - 4 \phi_0^\mathrm{in}(X(0))\right],
\end{equation}
and we have proved the following a-priori-bound:
\begin{Proposition}\label{boundf}
For all $t \in [0,T]$,
\[
\|f(t)\|_{\infty}\leq \|f^{\mathrm{in}}\|_\infty \exp\big[4\big(\|
\phi_{\mathrm{hom}}(t)\|_\infty+
\|\phi^{\mathrm{in}}_0\|_\infty\big)\big].
\]
\end{Proposition}
Notice that under the assumptions on the initial data for $\phi$ 
made above $\|\phi_{\mathrm{hom}}(t)\|_\infty \leq C (1+t)$ with $C$
depending on the data.
 
Combining this result with the one in 
\cite{CG} we obtain the following improvement on the continuation
criterion for classical solutions:
\begin{Proposition}\label{globalclassical}
Initial data as specified above launch a unique classical solution to the 
Cauchy problem 
for the Nordstr\"om-Vlasov system on a maximal time interval
$[0,T_{\mathrm{max}}[$, and if
\begin{equation}\label{momentumsupp}
\sup \{|p|:(x,p)\in\supp f(t),\ 0\leq t<T_{\mathrm{max}}\} < \infty
\end{equation}
then $T_\mathrm{max}=\infty$, i.e., the solution is global. 
\end{Proposition}
\noindent{\bf Proof:}
By \cite[Thm.~1]{CG} a unique solution exists and is global, 
provided $\phi$ remains bounded on $[0,T_{\mathrm{max}}[$ and the 
condition (\ref{momentumsupp}) holds.
But together with Proposition~\ref{boundf}, (\ref{momentumsupp})
implies that the source term $\mu$ in the field equation (\ref{wave})
is bounded. Hence a bound on $\phi$ follows from the
assumption on the momentum support, and the proposition
is established. \prfe

It is standard that the estimates above result in
a-priori-bounds for quantities like $\mu$ or $\rho$: 
\begin{eqnarray*}
\mu(t,x)
&=&
\int_{|p|\leq R}f(t,x,p)\,\frac{dp}{\sqrt{1+p^2}}+\int_{|p|\geq R}
f(t,x,p)\,\frac{dp}{\sqrt{1+p^2}}\\
&\leq& 
\|f(t)\|_{\infty}
\int_{|p|\leq R}\frac{dp}{\sqrt{1+p^2}}+R^{-2}\int \sqrt{1+p^2}f\,dp\\
&\leq& 
C R^2 \|f(t)\|_{\infty} + R^{-2}\int \sqrt{1+p^2}f\, dp = 
 C\left(\|f(t)\|_{\infty} \int \sqrt{1+p^2}f\,dp\right)^{1/2},
\end{eqnarray*}
where for the last step we choose
\[
R=\left(\int \sqrt{1+p^2}f\,dp / \|f(t)\|_{\infty}\right)^{1/4}.
\]
Squaring both sides of the estimate for $\mu$,
integrating in $x$, and using Proposition~\ref{boundf} and conservation
of energy implies the estimate
\begin{equation} \label{muest}
\|\mu(t)\|_2 \leq C e^{2 \|\phi_\mathrm{hom}(t)\|_{\infty}},\ t \in [0,T].
\end{equation}
Here $C$ denotes a positive constant which
depends only on $\|f^\mathrm{in}\|_\infty$, $\|\phi_0^\mathrm{in}\|_\infty$,
and the energy of the initial data, and which may change from line to line.
Similarly, using (\ref{distr}),
\begin{eqnarray*}
\rho(t,x)
&=&
e^{-\phi}\int_{|p|\leq R}f\,dp + e^{-\phi}
\int_{|p|\geq R}f\,dp\\
&\leq&
\|e^{-4 \phi_0^\mathrm{in}} f^\mathrm{in}\|_\infty e^{3 \phi} R^{3} +
e^{-\phi}R^{-1}\int \sqrt{1+p^2}\,f \,dp =
C \left(\int \sqrt{1+p^2}\,f \,dp\right)^{3/4},
\end{eqnarray*}
where we have chosen 
\[
R = e^{-\phi} \left(\int \sqrt{1+p^2}f\,dp\right)^{1/4}
\] 
in the last step. Elevating this estimate to the power $4/3$, integrating in $x$,
and observing conservation of energy we conclude that
\begin{equation} \label{rhojbound}
\|\rho (t)\|_{4/3},\ \|j (t)\|_{4/3}\leq C;
\end{equation} 
note that $\rho$ dominates $j$.

Combining (\ref{distr}) with Liouville's Theorem one can prove the following 
$L^q$ estimates on the distribution function. Since we will assume 
$L^\infty$ initial data for $f$ in the construction of weak solutions
Proposition~\ref{boundf} and the bound on the kinetic energy gives
us a bound on any $L^q$-norm of $f(t)$, and hence 
the following estimates are not used in the rest of the paper.
But they may be useful if one does not wish to consider bounded 
initial data for $f$.
\begin{Proposition}\label{Lqest}
For all $q\geq 1$, $\gamma\geq 3/q-4$, and $t\in [0,T]$ we have
\begin{equation}
\|e^{[3/q-4]\phi}f(t)\|_{q}
=
\|e^{[3/q-4]\phi_0^{\mathrm{in}}}
f^\mathrm{in}\|_{q}
\leq\|f^\mathrm{in}\|_q\exp\big(7\|\phi_0^\mathrm{in}\|_\infty\big),
\label{est1}
\end{equation}
\begin{equation}\label{est2}
\|e^{\gamma\phi}f(t)\|_{q}\leq \|f^{\mathrm{in}}\|_q\exp\Big[7
\big(\|\phi_{\mathrm{hom}}(t)\|_{\infty}+
\|\phi^{\mathrm{in}}_0\|_{\infty}\big)+
|\gamma|\,\|\phi_{\mathrm{hom}}(t)\|_{\infty}\Big].
\end{equation}
\end{Proposition}
\noindent{\bf Proof: }
For any smooth function $Q:\R\to\R$, 
Liouville's Theorem and (\ref{distr}) imply
\begin{equation}\label{liouville}
\int\!\!\!\int Q(f e^{-4\phi})\,e^{3\phi}\,dp\,dx=\mathrm{const};
\end{equation}
note that the $(x,p)$-divergence of the right hand
side of the characteristic system equals $- 3\, S\phi$ and so
\[
\det \frac{\partial(X,P)}{\partial(x,p)}(0,t,x,p) = 
\exp\left[3\,\phi(t,x) - 3\,\phi^\mathrm{in}(X(0,t,x,p))\right]. 
\]
With the choice $Q(z)=z^q$, (\ref{liouville}) implies (\ref{est1}). 
Moreover
\[
e^{[3/q-4]\phi}f = 
e^{\gamma\phi}e^{[3/q-4-\gamma]\phi_{\mathrm{hom}}}
e^{[3/q-4-\gamma]\psi}f . 
\]
Since $[3/q-4-\gamma]\,\psi\geq 0$ this yields 
\[
\|e^{[3/q-4]\phi}f(t)\|_q \geq  e^{-[3/q+4+|\gamma|]
\|\phi_{\mathrm{hom}}\|_{\infty}}\|e^{\gamma\phi}f\|_q,
\]
and (\ref{est2}) follows.\prfe

Although we have formulated the results of this section only for going forward in time
they hold equally well towards the past since the system is time reversible.

\section{Global weak solutions}
\setcounter{equation}{0}
The purpose of this section is to prove global existence 
of weak solutions to the Nordstr\"om-Vlasov system. 
We denote by $L^1_\mathrm{kin}(\R^6)$ the Banach space of the  
measurable functions $g:\R^{6}\to \R$ for which the norm 
\[
\|g\|_{1,\mathrm{kin}}=\int\!\!\!\int \sqrt{1+p^2}|g|\,dp\,dx
\]
is finite.

\smallskip

\noindent{\bf Theorem} {\em
For any triple $(f^\mathrm{in},\,\phi_0^\mathrm{in},\,\phi_1^\mathrm{in})$ 
such that for some $s>3/2$,
\[
0\leq f^\mathrm{in}\in L^1_{\mathrm{kin}}\cap L^\infty (\R^{6}),
\quad \phi_0^\mathrm{in}\in 
H^s(\R^3),\quad \phi_1^\mathrm{in}\in H^{s-1}(\R^3),
\]
there exists a global weak solution $(f,\phi)$ of the 
Nordstr\"om-Vlasov system, more precisely,
\[
f\in L^\infty(\R; L_{\mathrm{kin}}^1(\R^6))\cap L_{\mathrm{loc}}^\infty(\R; L^\infty (\R^{6})),
\quad 
\phi\in L^\infty_{\mathrm {loc}}(\R; H^1(\R^3)),
\]
with $\partial_t\phi,\, \nabla_x \phi\in L^\infty(\R; L^2(\R^3))$, 
$e^\phi\in H^1_{\mathrm{loc}}(\R\times\R^3)$, 
$f\geq 0$ a.~e., and the following holds:
\begin{itemize}
\item[(i)] $(f,\phi)$ solves (\ref{wave})--(\ref{vlasov}) in 
the sense of distributions.
\item[(ii)] The mapping 
\[
F:\R \to L^2(\R^{6})\times H^1(\R^3)
\times L^2(\R^3),\
t \mapsto (f(t),\phi(t),\partial_t\phi(t))
\]
is weakly continuous with 
$F(0)=(f^\mathrm{in},\,\phi_0^\mathrm{in},\,
\phi_1^\mathrm{in})$.
Moreover, for any $R>0$,
$\phi \in C(\R;L^2(B_R))$.
\item[(iii)] The energy at any time $t$
is bounded by its initial value, the local conservation law 
(\ref{localmasscons}) holds in the sense of distributions,
and the mass 
is conserved: 
\[
\int \rho(t)\, dx = \int \rho(0)\, dx \ \mbox{for a.~a.}\ t\in\R.
\] 
\end{itemize}
}

\smallskip

Here $B_R = \{ x \in \R^3 : |x| < R\}$, and $H^s(\R^3)$ denotes the usual 
Sobolev spaces. The proof proceeds in a number of steps:\\

\smallskip

\noindent
{\em Step 1: The regularized system.}\\
Let $0\leq \delta_n\in C^\infty_c(\R^3)$ be a mollifier satisfying the 
following conditions:
\[
\delta_n(x) = \delta_n(-x),\quad 
\int \delta_n(x)\, dx = 1,\quad
\supp\,\delta_n\subset B_{1/n},\quad n \in \N.
\]
The Nordstr\"om-Vlasov system is regularized by 
replacing the right hand side of (\ref{wave}) by $-\mu\star\delta_n\star\delta_n$, 
where $\star$ denotes convolution with respect to $x$. The regularized system is supplied
with regularized initial data 
\[
f_n^\mathrm{in},\ \phi_{0,n}^\mathrm{in}=g_n\star\delta_n,\
\phi_{1,n}^\mathrm{in} = h_n\star\delta_n,
\]
where
\[
0\leq f^\mathrm{in}_n\in C_c^\infty (\R^{6}),\ g_n,\,h_n\in 
C_c^\infty(\R^{3})
\] 
are such that $(f^\mathrm{in}_n)$ is bounded in $L^\infty(\R^6)$, and as $n\to\infty$,
\[
f^\mathrm{in}_n \to f^\mathrm{in} \ \mbox{in}\ 
L^1_{\mathrm{kin}}\cap L^2,\
g_n \to \phi_0^\mathrm{in}\ \mbox{in}\ H^s,\ 
h_n \to \phi_1^\mathrm{in}\ \mbox{in}\ H^{s-1}.
\]
The reason for regularizing $\mu$ and the data for $\phi$ in this particular way
will become obvious in the energy estimate below.
The regularized initial value problem has global smooth solutions. 
To see this observe that by (\ref{est2}), $||\mu(t)||_1$ is bounded. Hence
for any $n \in \N$, $\mu(t)\star\delta_n\star\delta_n$ is smooth and bounded in $L^\infty$,
together with all its derivatives, the bounds of course depending on $n$. 
Hence $\phi$ and $\partial_t \phi$ are bounded on any compact time interval,
together with all their spatial derivatives. 
It is then straight forward to see that a standard iteration
scheme like the one employed in \cite{CG} converges on any time 
interval to a smooth solution of the regularized initial value problem.
For more details we refer to the proof of the analogous result 
for the relativistic
Vlasov-Klein-Gordon system in  \cite[Sect.~3]{KRST}. 
Note that the regularized system remains time reversible so that
solutions really exist on all of $\R$ and not only on $[0,\infty[$.

In what follows, $(f_n,\phi_n)$ denotes the global smooth 
solution of the regularized initial value problem. It satisfies
the continuity equation
\[
\partial_t\rho_n+\nabla_x\cdot j_n = 0,
\]
where $\rho_n = e^{-\phi_n} \int f_n\,dp,\
j_n = e^{-\phi_n} \int \widehat{p}\,f_n\,dp$.

\smallskip
 
\noindent
{\em Step 2: Uniform bounds on $(f_n,\phi_n)$.}\\
First we observe that the energy of $(f_n,\phi_n)$ is not 
conserved, but
one can prove that it is bounded. In fact, a direct computation shows that
\begin{eqnarray*}
\frac{d}{dt}\Bigg\{\int\int\sqrt{1+p^2}f_n\,dp\,dx
&+&
\frac{1}{2}\int
\big(|\partial_t\phi_n|^2+|\nabla_x\phi_n|^2\big)\,dx\Bigg\}\\
&&
\qquad \qquad = \int\partial_t\phi_n(\mu_n-\mu_n\star\delta_n\star\delta_n)\,dx.
\end{eqnarray*}
Denote by $\tilde{\phi}_n$ the solution of the wave equation with 
the \textit{given} right hand side $-\mu_n\star \delta_n$ and initial data 
$\tilde{\phi}_{0,n}^\mathrm{in}=g_n$, 
$\tilde{\phi}_{1,n}^\mathrm{in}=h_n$. By uniqueness, 
$\phi_n = \tilde{\phi}_n\star \delta_n$. 
On the other
hand, the energy of $(f_n,\tilde{\phi}_n)$ is constant, and 
we conclude that
\begin{eqnarray}
\|f_n(t)\|_{1,\mathrm{kin}} + \frac{1}{2} \|\partial_t\phi_n(t)\|_2^2
&+&
\frac{1}{2} \|\nabla_x\phi_n(t)\|_2^2 \nonumber \\
&=&
\|f_n(t)\|_{1,\mathrm{kin}} + 
\frac{1}{2}\|\partial_t\tilde{\phi}_n (t)\star \delta_n\|_2^2 +
\frac{1}{2} \|\nabla_x\tilde{\phi}_n (t)\star \delta_n\|_2^2 \nonumber \\
&\leq&
\|f_n(t)\|_{1,\mathrm{kin}} + \frac{1}{2} \|\partial_t\tilde{\phi}_n(t)\|_2^2 +
\frac{1}{2} \|\nabla_x\tilde{\phi}_n(t)\|_2^2 \nonumber \\
&=& 
C; \label{energybound}
\end{eqnarray}
constants denoted by $C$ do not depend on $n$ or $t$.
The bound on $\partial_t \phi_n$ implies that 
$\|\phi_n(t)\|_2 \leq C_T$, for all $T>0$ and $t\in [-T,T]$,
where $C_T$ depends on $T$ but not on $n$.
Hence for any $T>0$,
\begin{equation} \label{phih1bound}
\|\phi_n(t)\|_{H^1} \leq C_T,\ t \in [-T,T].
\end{equation}

Let $\phi_{\mathrm{hom},n}$ 
denote the homogeneous part of the field in the modified
system. By Sobolev estimates for the solution of the homogeneous wave 
equation and the Sobolev embedding theorem it follows that 
\begin{equation} \label{phihombound}
\|\phi_{\mathrm{hom},n}(t)\|_\infty \leq C_T,\ t \in [-T,T].
\end{equation}
Hence the analogue of (\ref{phisplit}) and the fact that $\mu_n\star\delta_n\star\delta_n$ is 
non-negative imply that
\[
0 \leq e^{\phi_n (t)} \leq e^{\phi_{\mathrm{hom},n}(t)}
\leq C_T,\ t \in [-T,T].
\]
Hence
\[
\|\partial_t e^{\phi_n(t)}\|_2 \leq
\|e^{\phi_n}(t)\|_\infty \|\partial_t\phi_n(t)\|_2 \leq C_T,\ t \in [-T,T]
\]
and similarly for the $x$-derivatives. Hence for any $T>0$ and $R>0$
\begin{equation} \label{expphih1bound}
\|e^{\phi_n}\|_{H^1(]-T,T[ \times B_R)} \leq C_{T,R},\
\|e^{\phi_n(t)}\|_{H^1(B_R)} \leq C_{T,R},\ t \in [-T,T];
\end{equation}
clearly the same estimates hold for $e^{2\phi_n}$.
From Proposition~\ref{boundf}, (\ref{phihombound}), and the fact that 
the sequence
$(\phi_{0,n}^\mathrm{in})$ is uniformly bounded it follows that
\begin{equation} \label{fnbound}
e^{-4 \phi_n} f_n \leq C,\quad 0 \leq f_n,\leq C_T,\ t \in [-T,T]. 
\end{equation}
Hence by Eqn.\ (\ref{muest}) we conclude from
(\ref{fnbound}) and (\ref{energybound}) that 
\begin{equation}\label{munbound}
\|\mu_n(t)\|_{2} \leq C_T,\ t \in [-T,T].
\end{equation}
By (\ref{rhojbound}) we conclude that
\begin{equation} \label{rhojnbound}
\|\rho_n(t)\|_{4/3},\ \|j_n(t)\|_{4/3}\leq C,\ t \in \R.
\end{equation} 

\smallskip

\noindent
{\em Step 3: The weak limit.}\\
Here we use the estimates proved in Step 2 to obtain a weakly convergent
subsequence of $(f_n,\phi_n)$ whose limit will be a global weak solution.
The repeated extraction of suitable subsequences is not reflected in our 
notation. By
(\ref{energybound}), (\ref{phih1bound}), and (\ref{fnbound})--(\ref{rhojnbound}),
there exist 
\[
f\in L^\infty(\R;L^1_\mathrm{kin}(\R^{6})) \cap L^\infty_{\mathrm{loc}}(\R; L^\infty(\R^6)),\ 
\phi\in L_\mathrm{loc}^\infty(\R;H^1(\R^3))
\]
\[
\tilde{\mu}\in L^\infty_{\mathrm{loc}}(\R;L^2(\R^3)),\ 
\tilde{\rho},\tilde{j}\in L^\infty(\R;L^{4/3}(\R^3))
\]
such that
\[
f\geq 0\,\ \mbox{a.~e.},\ \partial_t\phi, \nabla_x \phi 
\in L^\infty(\R;L^2(\R^3)),
\]
and up a subsequence and for all $T>0$,
\begin{eqnarray*}
&&f_n\rightharpoonup f \textnormal{ in }L^2(]-T,T[\times\R^{6}),\\
&&\phi_n\rightharpoonup\phi\textnormal{ in }H^1(]-T,T[\times\R^3),\\
&&\partial_t\phi_n\rightharpoonup\partial_t\phi
\textnormal{ in }L^2(]-T,T[\times\R^3),\\
&&\mu_n\rightharpoonup\tilde{\mu}
\textnormal{ in }L^2(]-T,T[\times\R^3),\\
&&\rho_n,j_n\rightharpoonup\tilde{\rho},\tilde{j}
\textnormal{ in }L^{4/3}(]-T,T[\times\R^3),
\end{eqnarray*}
By a standard diagonal sequence argument we can choose the 
subsequence and its limit independent of $T>0$.
Next we observe that by (\ref{expphih1bound}) we can choose the subsequence 
such that $e^{\phi_n}$ converges weakly in $H^1(]-T,T[\times B_R)$ for any 
$T>0$ and $R>0$. Since that space is compactly embedded in 
$L^4(]-T,T[\times B_R)$ 
it follows that we can choose $\phi_n$ and $e^{\phi_n}$ to converge strongly in $L^4$, 
and by the Riesz-Fischer Theorem also pointwise a.~e. 
Hence
\[
e^{\phi_n}\rightharpoonup e^{\phi}\
\textnormal{in}\ H_\mathrm{loc}^{1}(\R \times \R^3),
\]
and the same is true for $e^{2\phi}$.
We shall prove now that $\tilde{\mu}=\mu,\,\tilde{\rho}=\rho$ 
and $\tilde{j}=j$ almost everywhere, where 
$\mu,\,\rho$ and $j$ are related to $(f,\phi)$ by (\ref{mudef}), 
(\ref{rhodef}), and (\ref{jdef}), respectively. 
As to $\tilde \mu$, we have for any $\chi\in C_c^\infty(\R\times\R^3)$
and any $R>0$,
\begin{eqnarray*}
&&
\int\!\!\!\int\left(\int f\,\frac{dp}{\sqrt{1+p^2}}-\tilde\mu\right)
\chi\,dx\,dt\\
&&
\quad=\int\!\!\!\int\left(\int_{|p|\leq R} f\,
\frac{dp}{\sqrt{1+p^2}}-\tilde\mu\right)\chi\,dx\,dt + 
\int\!\!\!\int\!\!\!\int_{|p|\geq
R}f\frac{dp}{\sqrt{1+p^2}}\chi\,dx\,dt\\
&&
\quad=\lim_{n\to\infty}\int\!\!\!\int\left(\int_{|p|\leq R} f_n\,
\frac{dp}{\sqrt{1+p^2}}-\mu_n\right)\chi\,dx\,dt + 
\int\!\!\!\int\!\!\!\int_{|p|\geq R}f\frac{dp}{\sqrt{1+p^2}}\chi\,dx\,dt\\
&&
\quad=\lim_{n\to\infty}\int\!\!\!\int\!\!\!\int_{|p|\geq  R} f_n\,
\frac{dp}{\sqrt{1+p^2}}\chi\,dx\,dt + \int\!\!\!\int\!\!\!\int_{|p|\geq
R}f\frac{dp}{\sqrt{1+p^2}}\chi\,dx\,dt.
\end{eqnarray*}
The last line can be estimated in modulus by
\[
\frac{C}{R^{2}} \left(\sup_{n\in \N} \|f_n(t)\|_{1,\mathrm{kin}}
+\|f(t)\|_{1,\mathrm{kin}}\right)\leq \frac{C}{R^2},
\]
where the constant $C$ is independent of $R$. 
Since $R>0$ is arbitrary we conclude that 
$\tilde \mu = \int f\,(1+p^2)^{-1/2}dp$ a.~e., and
\[
\mu = \int f \frac{dp}{\sqrt{1+p^2}} \in L_{\mathrm{loc}}^\infty(\R;L^2(\R^3)) \ \mbox{with}\
\mu_n\rightharpoonup\mu\textnormal{ in }L^{2}(]-T,T[\times\R^3)
\] 
for any $T>0$. Finally, if we define $\sigma_n = \int f_n dp$ so that
$\rho_n = e^{-\phi_n} \sigma_n$ then the sequence $(\sigma_n)$
is bounded in $L_{\mathrm{loc}}^\infty(\R;L^{4/3}(\R^3))$, and hence
as for $\mu$ we can show that 
$\sigma_n \rightharpoonup \int f\, dp$ in $L^{4/3}(]-T,T[\times\R^3)$.
On the other hand we have already seen that $e^{\phi_n} \to e^{\phi}$
strongly in $L^4(]-T,T[\times B_R)$. Hence
$\sigma_n = e^{\phi_n} \rho_n \to e^{\phi} \tilde\rho$ in the sense of 
distributions, and so $e^{\phi} \tilde\rho=\int f\, dp$ a.~e. Thus we have proved that
\[
\rho = e^{-\phi}\int f \, dp \in L^\infty(\R;L^{4/3}(\R^3))\ \mbox{with}\ 
\rho_n \rightharpoonup\rho \textnormal{ in }L^{4/3}(]-T,T[\times\R^3)
\]
for any $T>0$. It is obvious that the same assertion holds for $j$.

\smallskip

\noindent
{\bf Remark 1.} It is at this point that we need the $L^\infty$-bound
on $f$ and hence on the initial data: Since the fact that $e^\phi \in L^4_\mathrm{loc}(\R^4)$ 
seems optimal we must have
$\rho$ in the dual space $L^{4/3}$ which would not be true if
only an $L^q$-bound for $f$ with $q<\infty$, say $q=2$, were available.

\smallskip

The pair $(\phi,\mu)$ satisfies the inhomogeneous wave equation 
(\ref{wave}) in the sense of distributions. In fact
\begin{eqnarray*}
0 &=&
\int\!\!\!\int (\partial_t^2\phi_n-\bigtriangleup\phi_n+\mu_n\star\delta_n\star\delta_n)
\chi\,dx\,dt\\
&=&
\int\!\!\!\int (\phi_n\partial_t^2\chi
-\phi_n\bigtriangleup\chi+\mu_n(\delta_n\star\delta_n\star\chi))\,dx\,dt
\rightarrow
\int\!\!\!\int(\phi\partial_t^2\chi- \phi \bigtriangleup\chi +\mu\chi)\,dx\,dt
\end{eqnarray*}
for any test function $\chi \in C^\infty_c(\R \times \R^3)$.
The functions $\rho$ and $j$ satisfy the continuity 
equation (\ref{localmasscons}) in the sense of distributions. 
This implies quite easily that there exists {\em some}
constant to which the mass $\int \rho(t)\, dx$
is equal for almost all $t$, but we want to show that this constant
is equal to the initial mass the point being that a-priori we have
no continuity of the mass as a function of $t$.
Conservation of mass in the sense of the theorem is shown in Step~6 below.
The assertion on the total energy in item (iii) of the theorem is standard.

Passing to the limit in the linear part of 
the Vlasov equation (\ref{vlasov}) is not a problem. To complete the
proof that $(f,\phi)$ is a weak solution of (\ref{wave})--(\ref{vlasov}), 
we need to take care of the non-linear terms in the Vlasov
equation. This is done in the next step.

\smallskip

\noindent{\em Step 4: Momentum averaging.}\\
Let $R>0$ and $T>0$, $\psi\in C_c^\infty (\R^3)$ with
$\supp\psi\subset B_R$, and fix $\xi\in C_c^\infty(\R)$
such that $0\leq\xi\leq 1$ and $\xi=1$ on $[-T,T]$. Define
\[
\tilde{f}_n=\xi f_n,\quad g_0^n=(\xi'+\xi S\phi_n) f_n,\quad g_1^n=\xi F_n f_n,
\]
where $F_n=(S\phi_n)\,p+(1+p^2)^{-1/2}\nabla_x\phi_n$. 
Then we obtain
\[
S\tilde{f}_n = g_0^n+\nabla_p\cdot g_1^n,
\] 
cf.\ Eqn.\ (\ref{vlasovdiv}).
By the estimates of Step 3 the sequences $(g_0^n)$ and $(g_1^n)$ are
bounded in
$L^2(\R \times \R^3 \times B_R)$. Hence, by \cite{DL} the sequence
\[
\int \tilde{f}_n(\cdot,\cdot,p)\psi(p)\,dp
\]
is bounded in $H^{1/4} (\R\times\R^3)$, see also \cite{Gl,GLPS}.  
Since for all $R'>0$ and $T>0$, $H^{1/4}(]-T,T[\times B_{R'})$ 
is compactly embedded in $L^2(]-T,T[\times B_{R'})$, we conclude that
after extracting a subsequence, which, by a diagonal 
sequence argument, can be chosen independent of $T>0$ and $R'>0$,
\[
\int f_n(\cdot,\cdot,p)\,\psi(p)\,dp\to \int f(\cdot,\cdot,p)\,\psi(p)\,dp,
\quad \textnormal{strongly in }L^2(]-T,T[\times B_{R'}).
\]
Using this information we can prove that $(f,\phi)$ satisfies
the Vlasov equation in the form (\ref{vlasovdiv}). 
To see this let $\chi\in C_c^\infty(]-T,T[ \times B_{R'})$ and
$\psi \in C_c^\infty(B_{R})$. Then for the
non-linear part in (\ref{vlasovdiv}) we have
\begin{eqnarray*}
&&
\int\!\!\!\int\!\!\!\int
[\nabla_p\cdot(f_n F_n) + f_n S\phi_n ]\,\chi(t,x)\,\psi(p)\,dp\,dx\,dt\\
&&
\qquad =
\int\!\!\!\int
\biggl[
\int f_n  (- p \cdot \nabla_p\psi + \psi)\,dp \,\partial_t\phi_n\\
&&
\qquad \qquad \qquad - 
\int f_n  (p \cdot \nabla_p \psi \widehat{p} +
(1+p^2)^{-1/2} \nabla_p \psi - \psi \widehat{p})\, dp\,\cdot \nabla_x \phi_n 
\biggr]\,\chi\, dx\, dt \\
&&
\qquad \to
\int\!\!\!\int
\biggl[
\int f (- p \cdot \nabla_p\psi + \psi)\,dp \, \partial_t\phi \\
&&
\qquad \qquad \qquad - 
\int f \,(p \cdot \nabla_p \psi \widehat{p} +
(1+p^2)^{-1/2} \nabla_p \psi - \psi \widehat{p})\, dp 
\cdot \nabla_x \phi 
\biggr]\,\chi\, dx\, dt;
\end{eqnarray*}
note that the weight functions in the $p$ integrals of $f_n$
are compactly supported smooth functions as required by the momentum averaging 
argument. Hence the $p$ integrals converge strongly in 
$L^2(]-T,T[ \times B_{R'})$,
and since $\partial_t \phi_n$ and $\nabla_x \phi_n$
converge weakly in $L^2$ the assertion follows. 
By a standard density argument, we conclude that 
(\ref{vlasovdiv}) is satisfied in $\mathcal{D}'(\R\times\R^6)$.
This completes the proof of item $(i)$ of the theorem.

\smallskip

\noindent{\bf Remark 2.}
In the application of the momentum averaging argument the 
$L^\infty$-bound on $f$ could have been avoided, using the concept
of renormalization, cf.\ \cite{DL,Gl}. However, since we already needed
this bound above, cf.\ Remark 1, we prefer to avoid this
technical complication at this point.

\smallskip

\noindent{\em Step 5: Continuity in $t$.}\\
Using the Vlasov equation in the form (\ref{vlasovdiv}) for
the approximating sequence we have, for any test function
$\chi \in C^\infty_c(\R^6)$, 
\begin{eqnarray*}
&&
\int\!\!\!\int f_n(t)\, \chi\, dp\, dx
= \int\!\!\!\int f_n^{\mathrm{in}}\, \chi\, dp\, dx\\
&&
\qquad \quad +
\int_0^t
\int\!\!\!\int f_n(s) \Bigl[\widehat{p} \cdot \nabla_x \chi +
(S \phi_n)(s)\chi \\
&&
\qquad \qquad \qquad\qquad \qquad -
\left((S\phi_n)(s)p + 
(1+p^2)^{-1/2}\nabla_x \phi_n(s)\right)\cdot \nabla_p \chi
\Bigr]\,dp\,dx\,ds.
\end{eqnarray*}
The convergence of $f_n$ and $\phi_n$ is strong enough to pass to the
limit in the right hand side of this equation, and dropping $n$
in the right hand side we can use the resulting expression
to define a time dependent distribution 
$\tilde f (t) \in {\cal D}'(\R^6)$ which obviously is continuous in $t$
with respect to the usual topology of ${\cal D}'$, 
satisfies the initial condition
$\tilde f(0) = f^\mathrm{in}$, and  by construction coincides
for almost all $t$ with $f(t)$. By a density argument the
${\cal D}'$-continuity extends to 
continuity with respect to the weak topology of $L^2(\R^6)$.
A similar argument works for $\phi$ in which case the
continuous representative is defined by considering the first 
order formulation of the wave equation (\ref{wave}).

The stronger continuity assertion for $\phi$ follows from the 
Arzela-Ascoli theorem: Since $\partial_t \phi_n$ is bounded 
in $L^\infty(\R;L^2(\R^3))$ uniformly in $n$ the sequence
$(\phi_n)$ is equi-continuous as a sequence of $L^2(\R^3)$-valued
functions on $\R$. Moreover, for each $t\in \R$, $(\phi_n(t))$
is bounded in $H^1(\R^3)$ which is compactly embedded in $L^2(B_R)$.

\smallskip

\noindent
{\em Step 6: Conservation of mass.}\\
Since $\partial_t \rho_n + \div j_n = 0$ and $|j_n| \leq \rho_n$ we have 
for every $R>0$ and $t>0$,
\begin{eqnarray*}
\frac{d}{dt} \int_{|x| > R + t} \rho_n(t)\, dx 
&=&
- \int_{|x| = R + t} \rho_n(t)\, dS_x + \int_{|x| > R + t} 
\partial_t\rho_n(t)\, dx\\
&=&
- \int_{|x| = R + t} \rho_n(t)\, dS_x - \int_{|x| > R + t} \div j_n (t)\, dx\\
&=&
- \int_{|x| = R + t} (\rho_n(t) + \nu \cdot j_n (t))\, dS_x \leq 0
\end{eqnarray*}
where $\nu$ is the outer unit normal of the domain $\{|x| > R + t\}$. 
The analogous argument works for $t<0$ and the domain $\{|x| > R - t\}$. 
Hence 
\begin{equation} \label{finprop}
\int_{|x| > R + |t|} \rho_n (t)\, dx \leq \int_{|x| > R} \rho_n (0)\, dx,\ 
t\in \R,\ R>0,\ n\in \N.
\end{equation}
We claim that for almost all $t \in \R$,
\begin{equation} \label{massconsae}
\int \rho(t)\, dx =  \int\rho(0)\, dx.
\end{equation}
Let $\epsilon >0$ be arbitrary.
Since $\rho(0)$ is integrable, then we can choose $R>0$ 
such that 
\[
\int_{|x| > R} \rho(0)\, dx < \epsilon.
\]
By the convergence of the initial data
and (\ref{finprop}) we conclude that
\[
\int_{|x| > R + |t|} \rho_n(t)\, dx \leq \int_{|x| > R} \rho_n(0)\, dx 
< \epsilon
\]
for all $t\in \R$ and all sufficiently large $n\in \N$. 
Let $A \subset \R$
be measurable and bounded and denote by $\lambda(A)$ its Lebesgue measure. 
Then
\begin{eqnarray*}
\int_A \int \rho(t)\, dx\,dt
&\geq&
\int_A \int_{|x| \leq R + |t|} \rho(t)\, dx\,dt
=
\lim_{n \to \infty} \int_A \int_{|x| \leq R + |t|} \rho_n(t)\, dx\,dt\\
&=&
\lim_{n \to \infty} \int_A \left(\int \rho_n(t)\, dx - 
\int_{|x| > R + |t|} \rho_n(t)\, dx\right)\,dt\\
&>&
\lambda(A)\left( \int \rho(0)\, dx - \epsilon \right),
\end{eqnarray*}
and for sufficiently large $S>0$ we have by monotone convergence,
\begin{eqnarray*}
\int_A \int \rho(t)\, dx\,dt
&\leq&
\int_A \int_{|x| \leq S} \rho(t)\, dx\,dt + \lambda(A) \epsilon \\
&=&
\lim_{n \to \infty}
\int_A \int_{|x| \leq S} \rho_n(t)\, dx\,dt + \lambda(A) \epsilon \\
&\leq&
\lambda(A) \left( \int \rho(0)\, dx + \epsilon \right).
\end{eqnarray*}
Since $A$ was an arbitrary, bounded, measurable subset of $\R$ this implies
that there exists a set $M_\epsilon \subset \R$ of measure zero 
such that
\[
\int \rho(0)\, dx -  \epsilon \leq \int \rho(t)\, dx 
\leq \int \rho(0)\, dx + \epsilon,\ 
t \in \R \setminus M_\epsilon.
\]
Hence (\ref{massconsae}) holds on $\R \setminus \cup_{k\in \N}M_{1/k}$.

\smallskip

\noindent 
{\bf Remark 3.} The argument above makes use
of the relativistic nature of the system, i.e., of the finite
propagation speed of particles. We do not know if conservation of
mass in the sense of (\ref{massconsae}) holds without this property. 
In particular the above argument 
would also establish conservation of charge for the relativistic 
Vlasov-Maxwell system, but not for its 
\textit{non-relativistic} version in which velocity and momentum of 
the particles are equal. Note that the 
latter system is the one studied in detail in \cite{DL}. 

\smallskip

{\bf Acknowledgment:}
S.~C.\ acknowledges support by the European HYKE network 
(contract HPRN-CT-2002-00282).

\end{document}